\documentclass[fleqn,usenatbib]{mnras}

\usepackage{newtxtext,newtxmath}
\usepackage[T1]{fontenc}
\usepackage{ae,aecompl}
\usepackage{natbib}

\usepackage{graphicx}	% Including figure files
\usepackage{amsmath}	% Advanced maths commands
\usepackage{amssymb}	% Extra maths symbols

\title[Globular Clusters in UDGs]{Globular Clusters in Coma Cluster Ultra Diffuse Galaxies (UDGs): Evidence for Two Types of UDG?}

\author[Duncan A. Forbes et al.]{Duncan A. Forbes,$^{1}$\thanks{E-mail: dforbes@swin.edu.au}
Adebusola Alabi,$^{2}$
Aaron J. Romanowsky,$^{2,3}$
Jean P. Brodie$^{2}$
\newauthor
and Nobuo Arimoto$^{4,5}$
\\
$^{1}$ Centre for Astrophysics \& Supercomputing, Swinburne
University, Hawthorn VIC 3122, Australia\\
$^{2}$University of California Observatories, 1156 High Street, Santa Cruz, CA 95064, USA\\
$^{3}$Department of Physics and Astronomy, San Jos\'e State University, San Jose, CA 95192, USA\\
$^{4}$National Astronomical Observatory of Japan, Mitaka, Tokyo 181-8588, Japan \\
$^{5}$Astronomy Program, Department of Physics and Astronomy, Seoul National 
University, 599
Gwanak-ro, Gwanak-gu, Seoul, 151-742, Korea\\
}

\date{Accepted XXX. Received YYY; in original form ZZZ}

\pubyear{2019}

\begin{document}
\label{firstpage}
\pagerange{\pageref{firstpage}--\pageref{lastpage}}
\maketitle

\begin{abstract}

Ultra diffuse galaxies (UDGs) reveal extreme properties. 
Here we compile the largest study to date of 85 globular cluster (GC) systems around UDGs in the Coma cluster, using new deep ground-based imaging of the known UDGs and existing imaging from the Hubble Space Telescope of their GC systems. We find that the richness of GC systems in UDGs generally exceeds that found in normal dwarf galaxies of the same stellar mass.  These GC-rich UDGs imply 
halos more massive than expected from the standard stellar mass-halo mass relation. The presence of such overly massive halos presents a significant challenge to the latest  simulations of UDGs in cluster environments.
In some exceptional cases, the mass in the GC system is a significant fraction of the stellar content of the host galaxy. 
We find that rich GC systems tend to be hosted in UDGs of lower
luminosity, smaller size and fainter surface brightness. Similar trends are seen for normal dwarf galaxies in the Coma cluster. 
A toy model is presented in which the GC-rich UDGs are assumed to be `failed' galaxies within massive halos that have  largely old, metal-poor, alpha-element enhanced stellar populations. On the other hand,  GC-poor UDGs are more akin to normal, low surface brightness dwarfs that occupy less massive dark matter halos.
Additional data on the stellar populations of UDGs with GC systems will help to further refine and  test this simplistic model. 

\end{abstract}


%%%%%%%%%%%%%%%%%%%%%%%%%%%%%%%%%%%%%%%%%%%%%%%%%%

%%%%%%%%%%%%%%%%% BODY OF PAPER %%%%%%%%%%%%%%%%%%

\section{Introduction}

Since their discovery in 2015 \citep{2015ApJ...798L..45V},   UDGs have been the subject of much interest -- both observational and theoretical. As a working definition, 
UDGs have sizes R$_e$ $>$ 1.5 kpc and surface brightnesses $\mu_0$ $>$ 23.5 mag/sq. arcsec in the R band. They also have red colours distinguishing them from blue, star forming low surface brightness (LSB) galaxies that have been known for some years.  
The excitement around UDGs is largely due to the result that UDGs 
deviate strongly from normal galaxy scaling relations. In particular, they are not only `large' in size for their luminosity, 
but those measured are outliers in their dark matter content. Several UDGs are 
thought 
to be  `overmassive' containing $\sim 10\times$ more dark matter than normal dwarfs within their half-light radius  (\citet{2016ApJ...819L..20B};  \citet{2016ApJ...828L...6V}; \citet{2018ApJ...856L..31T}; \citet{2019MNRAS.484.3425M}). 
This finding would present a major theoretical puzzle 
since these UDGs are inferred to reside in near MW-mass halos, 
which in the standard picture of galaxy formation have {\it maximal} star formation efficiency, not {\it minimal}.

Evidence is mounting that UDGs may come in (at least) two different types. Key to this is the observation that some UDGs contain only a few globular clusters (GCs), while others are GC-rich  (\citet{2017ApJ...844L..11V}; \citet{2018MNRAS.475.4235A}; \citet{2018ApJ...862...82L}). 
As the mass of a GC system is known to be an indicator of halo mass, e.g.  \citet{2009MNRAS.392L...1S}, this suggests both normal and  massive halo UDGs exist.

These characteristics may be associated with `puffed-up dwarf' and `failed galaxy' scenarios, respectively (for a discussion of UDG origins see \citet{2019MNRAS.tmp.1490J}). 
Recent simulations such as \citet{2019MNRAS.486.2535D},
\citet{2019MNRAS.490.5182L}, 
%\citet{2019arXiv190406356L},  \citet{2019arXiv190805684T} and
\citet{2019arXiv190901347S} 
have successfully reproduced many of the properties of UDGs as puffed-up dwarfs. However, these simulations do not model the GC systems of UDGs nor do they predict any to have very massive halos. 
Although the non-detection of X-ray emitting halos around UDGs suggests that those with very massive halos may be in the minority  \citep{2019arXiv190605867K}, such UDGs present an important challenge to galaxy formation models that needs to be understood.
In the model of 
\citet{2016MNRAS.459L..51A},  UDGs are the high-spin
tail of otherwise normal dwarf galaxies. 
The non-rotation of the best studied UDG, DF44 in the Coma cluster by 
\citet{2019ApJ...880...91V}, suggests that this cannot be the full explanation.

To date, all published measurements of the dynamical mass of UDGs have been carried out on those UDGs with relatively large GC systems. Indeed such measures often use the motions of the GCs to determine the enclosed dynamical mass. 

Nearby `traditional' dwarf galaxies (i.e. dIrr, dE, dSph) also reveal a large range in the number of GCs per unit starlight, e.g.  \citet{2018MNRAS.tmp.2463F}; 
\citet{2010MNRAS.406.1967G}.  
Their host galaxies also have similar stellar masses to UDGs. However,   \citet{2018ApJ...862...82L} found that UDGs have, on average, higher 
GC specific frequencies (i.e. the number of GCs per unit galaxy luminosity) than normal early-type dwarf galaxies.
They also found a weak trend for higher GC specific frequencies 
%in larger and rounder UDGs, and 
in those UDGs located closer to the Coma cluster centre. They claimed to find trends between GC specific frequency and UDG size and surface brightness, although both these trends are extremely weak and need larger  samples to be confirmed. In their sample, 
\citet{2018MNRAS.475.4235A} 
identified nine Coma cluster UDGs to be particularly GC-rich. Contrary to the findings of \citet{2018ApJ...862...82L} 
they found these GC-rich UDGs to have, on average, smaller sizes. They did not examine trends with surface brightness or cluster-centric distance. 
%Neither study examined whether GC richness varied with with host galaxy colour.  

%An interesting possible explanation for the GC systems of UDGs is that they are the surviving population of GCs, while a significant fraction of the UDG stellar mass is composed of disrupted GCs \citep{2016ApJ...822L..31P}. We note that the stellar population properties of Coma UDGs studied to date indicates that the stellar mass is not dominated by GC-like stellar populations \citep{2018MNRAS.479.4891F}. 

Here we combine the GC studies of Coma cluster UDGs by 
 \citet{2018ApJ...862...82L} and 
\citet{2018MNRAS.475.4235A} with that of \citet{2017ApJ...844L..11V}, and new deep  photometry of their host galaxies by Alabi et al. (2019, in prep). 
We include the galaxies with sizes 0.7 $<$ R$_e$ $<$ 1.5 kpc from 
\citet{2018MNRAS.475.4235A} 
to probe the interface between {\it bona fide} UDGs (with R$_e$ $>$ 1.5 kpc) and their smaller counterparts.  Hereafter, we refer to these `small UDGs' as low surface brightness (LSB) dwarfs.
This gives us the largest sample to date of UDGs with with measured GC counts  (mostly from imaging with HST/ACS). We look for trends between GC systems and their host galaxies. In particular, we examine galaxy stellar and halo mass, 
absolute magnitude, size, colour and surface brightness. We also present a toy model in which UDG  properties range from GC-rich failed galaxies to GC-poor puffed-up dwarf  galaxies. 
In this work we assume a distance of 100 Mpc to the Coma cluster.
%(and a corresponding angular distance of 95.54 Mpc). 

\section{UDG Host Galaxy Properties}

Low surface brightness galaxies in the Coma cluster were cataloged by \citet{2016ApJS..225...11Y}. This catalog was created from deep Subaru imaging in the R band. They selected galaxies with half-light radii (R$_e$) greater than 0.7 kpc and mean surface brightnesses within the half-light radius of 24 $<$ $\mu_e$ $<$ 27 R mag. per sq. arcsec. 
%They measured total R magnitudes with a repeatability of $\le$ 0.08 mag. 

Traditionally, GC systems have been compared to the total V band luminosity of their host galaxy (a quantity called the GC specific frequency). Although V band magnitudes exist in the literature for some Coma UDGs, a homogeneous V band photometric study has not been available until now.

Using new deep V band Subaru imaging 
covering 4.2 sq. degrees of the Coma cluster,  Alabi et al. (2019, in prep) identifies galaxies in common with the R band catalog of  \citet{2016ApJS..225...11Y} supplemented by those from the \citet{2015ApJ...798L..45V} Dragonfly catalog. The new catalog reaches a depth of 27.1 mag per sq. arcsec in the R band and detects several new UDG candidates. All UDGs meet the UDG selection criteria, e.g R$_e$ $>$ 1.5 kpc and $\mu_0$ $>$ 23.5 R mag sq. arcsec. Alabi et al. fits a single Sersic profile to each galaxy in both the V and R bands, measuring their R$_e$, Sersic index (n), total magnitude, mean surface brightness within 1~R$_e$ ($<\mu>_e$) and V--R colour within a matched 1~R$_e$ aperture. Measurements of the half-light radii, R band magnitudes and surface brightnesses show good agreement with the Yagi catalog (Alabi et al. 2019, in prep).
%although we note that total magnitudes are systematically fainter than those of \citet{2017ApJ...844L..11V}. 
The measurement uncertainties, based on repeated measurements of galaxies that appear in multiple frames, are around $\pm$ 0.13 kpc in half-light radius, $\pm$ 0.21 mag. per sq. arcsec in surface brightness, $\pm$ 0.19 mag in magnitude, 
%$\pm$ 0.18 mag in R band magnitude 
and $\pm$ 0.24 mag in V--R colour for the UDGs. In this work we use preliminary  photometry from Alabi et al. (2019, in prep.) but we do not expect the final published photometry to affect our conclusions.

%The UDG properties we use in this study come from the new measurements by Alabi et al.. 

%We note that 
%nine UDGs in the study of \citet{2018ApJ...862...82L} and five from
%\citet{2018MNRAS.475.4235A} 
%fall outside of the Alabi et al. field-of-view. 
%For these galaxies we have taken their half-light radii, total V band magnitudes and mean surface brightnesses in the V band directly from their work.
%We calculated the galaxy stellar mass assuming a fixed mean colour of V--R = 0.52.
%Similarly, for 10 UDGs in the \citet{2018MNRAS.475.4235A} 
%study, we use half-light radii from their work with total magnitudes and mean surface brightnesses from the original \citet{2016ApJS..225...11Y} catalog assuming V--R = 0.52. ***

We exclude UDGs known to lie outside of the Coma cluster, i.e. DF3 (Yagi214)  \citep{2017ApJ...838L..21K} and Yagi771 
\citep{2018MNRAS.479.3308A}.
We also note DF42, DF44 and DFX2 may belong to a small group of galaxies currently infalling into the Coma cluster (van Dokkum et al. 2019) but we keep them in this study.

As previously noted, the catalog of \cite{2016ApJS..225...11Y} and the GC study of \citet{2018MNRAS.475.4235A} included galaxies with sizes smaller than the traditional definition of a UDG (i.e. 0.7 $<$ R$_e$ $<$ 1.5 kpc). These `small UDGs' provide a useful comparison with {\it bona fide} UDGs and we refer to them as LSB dwarfs.

\begin{figure}
	\includegraphics[width=7.1cm, angle=-90]{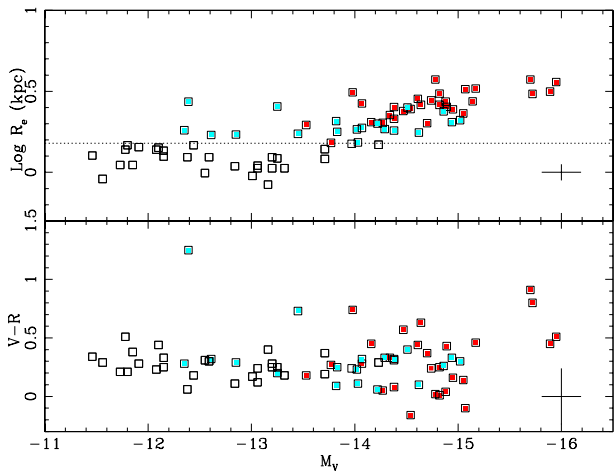}
    \caption{Size and colour-magnitude relations for Coma UDGs with observed GC systems. Red symbols are UDGs with Dragonfly IDs from the sample of  \citet{2017ApJ...844L..11V} and cyan symbols are non-Dragonfly UDGs from  \citet{2016ApJS..225...11Y}. 
    %One non-Dragonfly UDG with a colour V--R = 1.25 is excluded from the plot. 
    Open symbols are LSB dwarfs  with sizes 0.7 $<$ R$_e$ $<$ 1.5 kpc.
    %Measurements of half-light radius (R$_e$), V--R colour and absolute magnitude (M$_V$) are new measurements from Alabi et al. (2019, in prep.). 
    Typical measurement uncertainties are shown in the lower right of each panel.
    The traditional size limit for UDGs is R$_e$ $>$ 1.5 kpc (shown by a dotted line). %Both UDG size and V--R colour show only a weak trend with magnitude. 
%    A strong size-magnitude trend is present in the data (with a best-fit shown by a solid line), while the colour-magnitude relation is essentially flat.
    }
    
    \label{fig:cmr}
\end{figure}

In Fig.~\ref{fig:cmr} we show the size--magnitude and colour--magnitude relations for our sample of Coma UDGs with observed GC systems (see section 4 for details). We also show, with different symbols, the Coma LSB dwarfs. %with 0.7 $<$ R$_e$ $<$ 1.5 kpc. 
The V--R colour of UDGs reveals only a weak trend with magnitude.  This is partly a consequence of the limited sensitivity of V--R colour to metallicity and the small luminosity range of UDGs. 
We note that 
\citet{2015ApJ...798L..45V} measured a mean g--i colour 
of 0.8 $\pm$ 0.1 for their sample of Dragonfly UDGs. 
The trend of size with magnitude is partly due to surface brightness limits (which form diagonal lines in a size-magnitude plot).  Here the bright surface brightness limit is set by the definition of UDGs, whereas a lower limit to surface brightness is set by the depth of the catalog used.
The Coma LSB dwarfs tend to be less luminous on average than {\it bona fide} UDGs but with similar V--R colours. 

% The plot also shows a best-fit relation to the full sample of:\\

%log R$_e = -0.126(\pm0.015)~M_V - 1.49(\pm0.21)$ \\

%A detailed discussion of the size-magnitude relation for Coma UDGs, from CFHT g and i band imaging, can be found in Daniele \& van Dokkum (2019). 
%As can be seen in Fig.~\ref{fig:cmr},  there are two extremely blue UDGs (DF40 and DF41). 
%with V--R = --0.05 and  DF40 with V--R = 0.16. 

%A CFHT image of the Coma cluster in the g and i bands, Danieli et al. (2019) gives total magnitudes and g--i colours but only for a subset of Yagi catalog objects. **

\section{UDG Stellar Masses}

%Based on the absolute V band magnitude we estimate UDG stellar masses. 
%Ideally one would use age and metallicity information to calculate the appropriate M/L ratio. However, a
Only about a dozen higher surface brightness Coma UDGs have measured spectroscopic stellar population properties (  \citet{2018MNRAS.479.4891F};  \cite{2018MNRAS.478.2034R}; \citet{2018ApJ...859...37G};
\citet{2019ApJ...884...79C}).
%\citet{2019arXiv190105489C}). 
These studies generally found intermediate-to-old ages and low metallicities. The V band mass-to-light ratio for such stellar populations is around M/L$_V$ = 1 for a Kroupa IMF.
The IMF in UDGs is currently unconstrained; using a Salpeter IMF  would systematically increase the derived masses by $\sim$0.2 dex. Here we convert M$_V$ for UDGs (and LSB dwarfs) into stellar masses assuming 
M/L$_V$ = 1 (while noting there may be  a factor of two variation in this quantity for individual galaxies) and a V band solar luminosity of +4.83. 

\begin{figure}
	\includegraphics[width=7.1cm, angle=-90]{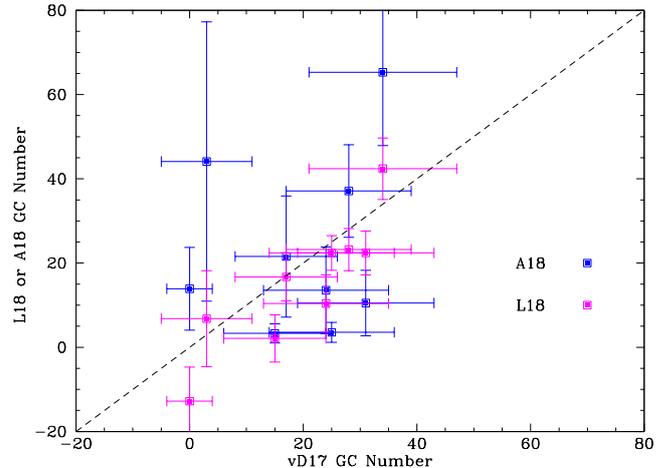}
    \caption{Comparison of GC numbers from different studies. The plot shows the studies of A18 =  \citet{2018MNRAS.475.4235A} and L18 =  \citet{2018ApJ...862...82L}  to those of vD17 =  
    \citet{2017ApJ...844L..11V}
    for UDGs in common. The dashed line is a 1:1 relation. Within the errors, the three studies are in reasonable agreement. 
}
    
    \label{fig:common}
\end{figure}

\section{UDG GC Systems}

In this work we use three literature studies of GC systems around Coma cluster UDGs.
%In each of these works, the imaging %of low surface brightness (LSB) %galaxies from the 
%\cite{2016ApJS..225...11Y}
In each of these works, globular cluster detection is carried out using HST/ACS imaging, with the majority from the Coma Treasury Survey \citep{2008ApJS..176..424C}. 
Primarily, we take GC counts from the study of  \citet{2017ApJ...844L..11V}. For objects not covered by van Dokkum et al., we take GC counts from \citet{2018ApJ...862...82L}. Finally, this is supplemented by GC counts from \cite{2018MNRAS.475.4235A}. 
The former two studies used the original size definition of a UDG, i.e. $R_e >$ 1.5 kpc. 
Here we supplement this sample with LSB dwarfs with sizes 0.7 $< $ R$_e$ $<$ 1.5 kpc from 
\citet{2018MNRAS.475.4235A}. 
%The former two studies used a conventional approach of subtracting a statistical background from the GC candidate counts around each UDG. 
We note that the \citet{2018MNRAS.475.4235A} study used a mixture model to perform a statistical correction and estimate the 10th, 50th and 90th percentiles of GC counts associated with each galaxy. Here we assume that their posteriors have a Gaussian distribution  to
estimate mean values and 1$\sigma$ errors, similar to those quoted by the two other studies. 
%For example, Yagi534 with 10, 50 and 90th percentiles of 23, 35 and 51 translate into $N_{GC}$ = 37 $\pm$ 11. 

%, and note that only two dozen of the sample have confirmed redshifts associated with the Coma cluster \citep{2018MNRAS.479.3308A}. 

%We note Yagi419, which is absent from the \citet{2018ApJ...862...82L} study (as they suggested it may be a superposition of two galaxies), is consistent between \citet{2017ApJ...844L..11V} of $N_{GC}$ = 23 $\pm$ 11 and \citet{2018MNRAS.475.4235A} of $N_{GC}$ = 13.5 $\pm$ 9.6. We include this in our sample as a single galaxy. 

In Fig.~\ref{fig:common} we compare the results of GC counts in UDGs from \citet{2018ApJ...862...82L} and \cite{2018MNRAS.475.4235A} with the study of 
\citet{2017ApJ...844L..11V}.
Both the \citet{2018ApJ...862...82L} and \citet{2018MNRAS.475.4235A} samples show reasonable agreement within the errors (see also figure 2 of \citet{2018ApJ...862...82L}). 
%, with the main exception being Yagi122 which  \cite{2018ApJ...862...82L} suggests is devoid of GCs. 
The rms spread in the data between the different studies is around $\pm$15 in GC number. 

As noted above, we take GC measurements in the following priority order: 
\citet{2017ApJ...844L..11V}
\citet{2018ApJ...862...82L}, and finally \cite{2018MNRAS.475.4235A}. 
This gives a sample of 76 Coma cluster UDGs and LSBs (UDGs with sizes 0.7 $<$ R$_e$ $<$ 1.5 kpc) with GC measurements.
Nine UDGs from the study of \citet{2018ApJ...862...82L} lie outside of the 
field-of-view of the Subaru imaging used by 
Alabi et al. (2019). For these nine we take GC counts, M$_V$, R$_e$ and V band $<\mu>_e$ directly from 
\citet{2018ApJ...862...82L}; colours are not available.  We assume M/L$_V$ = 1 to calculate host galaxy stellar masses and V--R = 0.3 to convert from V to R band surface brightness. This gives a total sample of 85 galaxies with GC counts -- the largest sample studied to date. 
We note that several galaxies in the sample are deemed to have no GCs (i.e. they are consistent with zero GC counts after a statistical background subtraction). 
In the Appendices we list the GC number counts for our combined sample of Coma cluster UDGs, i.e. Table A1 for Dragonfly IDs  \citep{2017ApJ...844L..11V}, Table A2 for Yagi only IDs  \citep{2016ApJS..225...11Y} and the additional 9 UDGs from \citet{2018ApJ...862...82L} in Table A3.

\section{Dwarf Galaxy Comparison Samples}

To help interpret the GC systems of UDGs we employ several comparison samples. As well as the LSB dwarfs mentioned above, we include local dwarf galaxies and Coma early-type dwarf galaxies with observed GC systems.
Local dwarfs are relevant as UDGs may have formed their GCs in low density environments before falling into clusters, e.g. \citet{2018MNRAS.479.3308A}. 
%UDGs have been proposed to be simply puffed-up dwarf galaxies (\citet{2018MNRAS.475.4235A}). 
If UDGs are simply puffed-up dwarfs, then it is relevant to compare them to a sample of Coma early-type dwarf galaxies. 

The local dwarf sample consists of GC systems for 14 Local Group dwarfs (including a re-assembled Sgr dwarf galaxy) and 29 Local Volume dwarfs in low density environments from \citet{2018MNRAS.tmp.2463F}.
This therefore includes early (dE and dSph) and late-type (dIrr and Im) dwarfs. Many of them are satellites of a more massive galaxy, which may have modified their dark matter halos and GC systems.
For each galaxy we take the quoted number of old GCs, the total mass of the GC system (which is the sum of the individual masses), the host galaxy absolute magnitude and the host galaxy stellar mass. 
%(from \citet{2012AJ....144....4M}. 

The other comparison sample is based on the 
study of \citet{2018ApJ...862...82L} who measured the GC content of 50  early-type dwarfs in the core of the Coma cluster from HST/ACS imaging. As they reside mostly in the core, such dwarfs tend to be older and redder than typical Coma dwarfs \citep{2009MNRAS.392.1265S}. 
Here we take the GC counts, host galaxy absolute magnitudes from Lim et al. and again assume M/L$_V$ = 1 to derive host galaxy stellar masses.

%\citet{2018ApJ...862...82L} took the I band magnitudes from \citet{2014MNRAS.445.2385D} 
%and assumed V--I = 0.4. This colour is too blue, e.g. \citet{2004AJ....128.2797V} find a mean colour of V--I = 1.02 $\pm$ 0.03 for Virgo dE galaxies. Here we assume V--I = 1.0 for each Coma dE galaxy and use this to 
%calculate revised M$_V$ and stellar masses from the original I band magnitude. 
%We use log(M/L$_V$) = 1.826(V--I) -- 1.629 from \citet{2013MNRAS.430.2715I} and 
%assume M$_V$ = +4.83 for the Sun.
%The resulting absolute magnitudes and stellar masses of both dwarf samples is comparable to that of the Coma UDGs.
%Our new M$_V$ values result in systematically lower GC specific frequency values for the Coma dwarfs compared to Lim et al. 

%Again these dwarf galaxies have similar stellar masses to the Coma UDGs. We take the number of GCs and host galaxy M$_V$, however, we assume a more realistic value of V--I = 1.0 than V--I = 0.4 for these dwarfs  \citep{2018ApJ...862...82L}. 

\section{Results}

In the subsections 6.1--6.3 we examine trends between the stellar mass of the host UDG and the number of associated GCs including the directly related quantities of GC system mass and galaxy halo mass. In the latter subsections 6.4--6.7 we normalise the GC counts by the V band luminosity of the host galaxy (called GC specific frequency, S$_N$).  Given that we apply a constant mass-to-light ratio this is equivalent to  normalising by stellar mass. We examine trends between S$_N$ and host galaxy properties. 
We also include other dwarf galaxy comparison samples when data is available.

\subsection{GC System Numbers}

We begin by examining the number of GCs versus host galaxy stellar mass for UDGs, and contrast them with Coma and local dwarf galaxies. 

%for stellar masses of $5 \times 10^6 - 10^8 M_{\odot}$, although there some dwarfs with single GCs that exhibit a much higher GC-to-stellar mass ratio (e.g. Eridanus II, with a stellar mass of $\sim$ $6 \times 10^4 M_{\odot}$, has 1 GC). 

\begin{figure}
	\includegraphics[width=7.1cm, angle=-90]{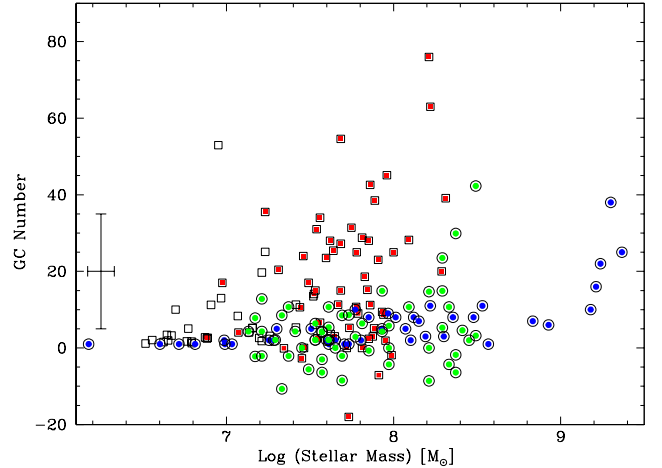}
    \caption{Number of globular clusters versus host galaxy stellar mass. Coma UDGs are shown by red squares, Coma LSB dwarfs by open squares, Coma early-type dwarfs by green circles and local dwarfs by blue circles.  Some Coma galaxies have no GCs after background subtraction. For local dwarfs only those with GC systems are shown.
    The typical uncertainty for Coma UDGs is shown on the left hand side. The UDG with the highest GC number (76) is DF44.  
    %Dashed lines represent 1 GC per log stellar mass of 6.7 and 8.0. Galaxies with 1 GC at the low and high mass extremes are Eridanus II and the SMC.
}
    
    \label{fig:num}
\end{figure}

Fig.~\ref{fig:num} shows that 
although some UDGs overlap with the other dwarf galaxy samples, many UDGs reveal considerably more GCs than Coma dwarfs or local dwarfs. Indeed, over half of the UDGs have more GCs per unit galaxy mass than those of the highest dwarf galaxies. The UDG with the highest GC number (76) is DF44 \citep{2017ApJ...844L..11V}. We also note that the three Coma early-type dwarfs with particularly rich GC systems also have large error bars, i.e. $N_{GC}$ = 23.5 $\pm$ 11.6, 29.9 $\pm$ 14.9 and 42.3 $\pm$ 13.6. Some of the LSB dwarfs have high GC counts but only the very highest mass local dwarfs with log stellar mass $>$ 9 reveal high GC counts, e.g. NGC~1427A with its 38 reported GCs. 

Half a dozen UDGs have no discernible GC system after a statistical background subtraction. Such systems have zero or negative GC counts quoted in the original studies. The host galaxies of such GC-free systems cover the full range of UDG stellar masses. In the subsequent analysis, we found no clear galaxy property that correlated with the lack of GCs and we do not consider them further.
%The same is true for several Coma early-type dwarfs. 
For the local dwarf comparison sample, we only include populated GC systems, which range from 1 GC per host galaxy to several tens of GCs for the highest mass dwarfs, and do not consider GC-free galaxies. 

%In terms of a mean GC number and 1$\sigma$ dispersion we find ***
%for the UDGs, ** for the early-type dwarfs and ** for the local dwarfs. 

\subsection{GC System Masses}

It is also interesting to compare the total mass in a GC system to the stellar mass of the host galaxy. 
Unfortunately, little is known about the luminosity (or mass) function of the GC systems associated with UDGs. For the GC-rich UDGs DF44, DFX1 \citep{2017ApJ...844L..11V} and DF17 \citep{2016ApJ...822L..31P} 
the bright end of the GC luminosity function appears to be similar to that of the `universal' GCLF \citep{2017ApJ...844L..11V}.  These UDGs may have GC systems similar to those of the Milky Way and local giant galaxies. 
%The GC luminosity function of a GC-poor UDG was recently studied by \citet{2019MNRAS.486..823R}. 
If the typical GC luminosity is M$_V$ $\sim$ --7.4 then the typical mean GC mass is $\sim 10^5 M_{\odot}$. Here we use this as the typical mean mass for a GC in all UDGs but note that the actual mean GC mass is unknown. We also adopt this mean GC mass for Coma LSB dwarfs and Coma early-type galaxies.
%The mean GC mass in low luminosity galaxies has been found to be less than that for high luminosity galaxies \citep{2017ApJ...836...67H} 
%and we adopt  10$^5$ M$_{\odot}$ for the mean GC mass for the \citet{2018ApJ...862...82L} Coma early-type dwarfs. 
Local dwarf galaxies reveal a wide range in their mean GC mass. 
Here we take the mass of the GC system directly from  \citet{2018MNRAS.tmp.2463F}, who summed the observed individual GC masses. 
We note that some GC-poor local dwarfs have a mean GC mass closer to $10^4 M_{\odot}$.

\begin{figure}
	\includegraphics[width=7.1cm, angle=-90]{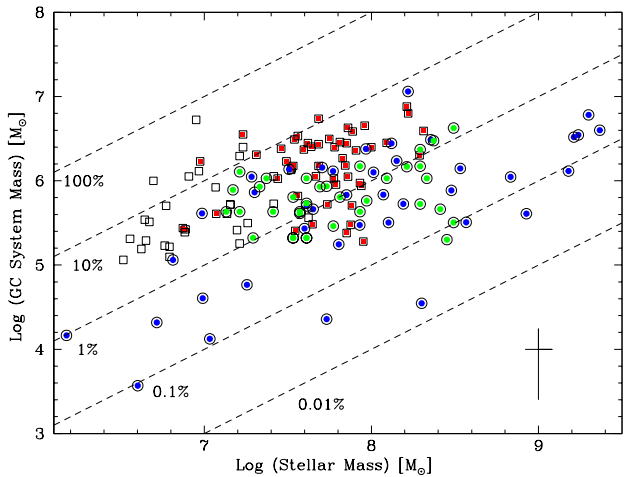}
    \caption{Mass in globular cluster system versus host galaxy stellar mass. Coma UDGs are shown by red squares, Coma LSB dwarfs by open squares, Coma early-type dwarfs by green circles and local dwarfs by blue circles. We assume a mean GC mass of $10^5 M_{\odot}$ for the UDGs and the Coma dwarfs. For the local dwarfs individual GC masses are summed. Dashed lines represent GC system to galaxy stellar mass ratios. The typical uncertainty for Coma UDGs is shown on the right hand side. Some UDGs (and LSBs) have a GC system mass that is a significant fraction of the  host galaxy stellar mass. 
}
    \label{fig:mstar}
\end{figure}

In Fig.~\ref{fig:mstar} we show the total GC system mass vs total stellar mass of the host galaxy. As well as Coma UDGs we include the Coma LSB and early-type dwarfs, and the local dwarfs samples. 
Each sample reveals a wide range of GC system mass-to-galaxy stellar mass ratios, from a fraction of a percent to around 10\%. On average  UDGs reveal higher GC system mass-to-galaxy mass ratios than early-type or local dwarfs, and at times the GC system mass approaches a significant fraction of  the total stellar mass of the host UDG. We remind the reader that this diagram depends on the assumption that the mean GC mass in UDGs, LSB dwarfs and early-type dwarfs has the `universal value' of $10^5 M_{\odot}$, but also note that similar general trends are seen when simply GC number is used (see Fig.~\ref{fig:num}).
 %, which does not depend on an assumption of the mean GC mass. 

\subsection{Halo Masses}

\begin{figure}
	\includegraphics[width=7.1cm, angle=-90]{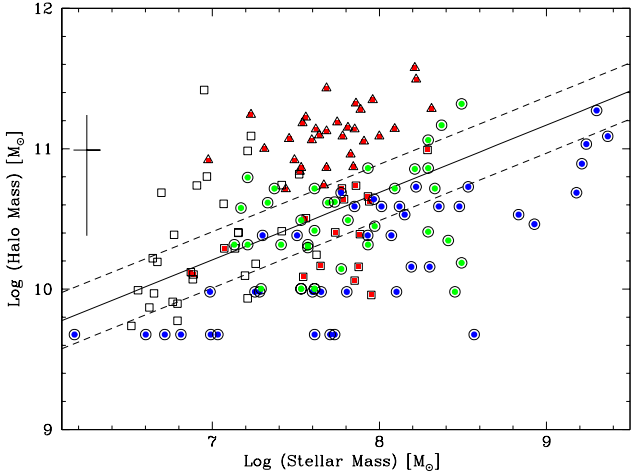}
    \caption{Halo mass vs stellar mass. The halo mass is derived from the scaling between GC system number and halo mass (see text for details). 
    The solid line shows the extrapolated stellar mass--halo mass relation for normal galaxies of \citet{2017MNRAS.470..651R}. The dashed lines show the typical scatter of $\pm$ 0.2 dex found at higher masses.
    Coma UDGs are shown by red squares and red triangles (if they lie above the relation upper limit). 
    Coma LSB dwarfs are shown by open squares, Coma early-type dwarfs by green circles and local dwarfs by blue circles. The typical uncertainty for Coma UDGs is shown on the left hand side. Some 65\% of Coma UDGs lie above the upper limit of the relation to higher halo masses. 
}
    \label{fig:halo}
\end{figure}

Near-linear empirical relations have been found between the total halo mass of the host galaxy and both the number of GCs \citep{2019arXiv190100900B} and 
the total mass of the GC system \citep{2009MNRAS.392L...1S} for normal galaxies. Perhaps the best studied Coma UDG is that of DF44. As shown in \citet{2019arXiv190100900B}, its halo mass based on radially extended kinematics and its halo mass based on its GC count are in excellent agreement. 
%{\bf We note that the Virgo UDG VCC1287 (Gannon et al. 2019, in prep) also appears to be consistent with the Burkert \& Forbes relation. 
Nevertheless, we emphasise to the reader that here we make the assumption that this relation is valid for all Coma galaxies, including UDGs. Thus we use eq. 1 of \citet{2019arXiv190100900B} to convert our GC number counts into total (pre-infall) halo masses for UDGs and LSB dwarfs (and noting that the relation begins to break down for log halo masses $<$ 10). 
The resulting halo masses are shown in Fig.~\ref{fig:halo}. We also show the stellar mass--halo mass (SMHM) relation of \citet{2017MNRAS.470..651R} extrapolated to lower masses after correcting for the sign error in eq. 66 (Rodriguez-Puebla, priv. comm).
%Mh = 6.848 + 0.48logM*
%eg at M*=10^10, Mh = 11.65
This work employed a semi-empirical approach to subhalo abundance matching. It agrees well with weak lensing apporoaches. 
The scatter about this relation at  L$^{\ast}$ is $\sim$0.2 dex but is somewhat larger at the lower stellar masses typical of UDGs. 
Both the Coma LSB and early-type dwarfs tend to scatter about the relation while the local dwarfs tend to lie below with lower halo masses (many are satellites and may experience tidal stripping of their halos).  
For the UDGs, 65\% have systematically 
higher halo masses than expected compared to the upper limit of the relation. This may point to two types of UDG, i.e. those with normal (dwarf-like) halo masses and those with abnormally massive halos. The latter being rich in GCs. 
%These two types of UDG being equivalent to poor and rich in GC counts respectively.

The tendency for many Coma UDGs to have higher halo masses than the SMHM relation is in stark contrast to the latest simulations for UDGs. In \citet{2019arXiv190805684T}
UDGs in a Virgo cluster-like potential lie on the SMHM relation prior to infall. And after infall they scatter to {\it lower} halo masses due to stripping of their dark matter halos. Similarly, in the Auriga simulations of Milky Way-like halos by  \citet{2019arXiv190406356L}, {\it } all of their UDGs have dwarf-like halos with none being failed masive halo galaxies. 
Their highest log halo mass is 10.5. 
While these simulations appear to successfully reproduce many properties of UDGs as puffed-up dwarfs, they fail to reproduce the population of massive halo UDGs. GCs may hold the key, so that a full understanding of all UDGs may not be available until realistic GC systems are included in simulations.

\subsection{GC Specific Frequency  vs Absolute Magnitude}

The traditional method of comparing the GC system richness of different galaxies is via GC specific frequency which normalises the GC number by the host galaxy luminosity in the V band, i.e. S$_N = N~10^{0.4(M_V + 15)}$. \citet{2018ApJ...862...82L} showed that Coma UDGs have, on average, higher S$_N$ values than dwarf galaxies of comparable host galaxy luminosity. 

\begin{figure}
	\includegraphics[width=7.1cm, angle=-90]{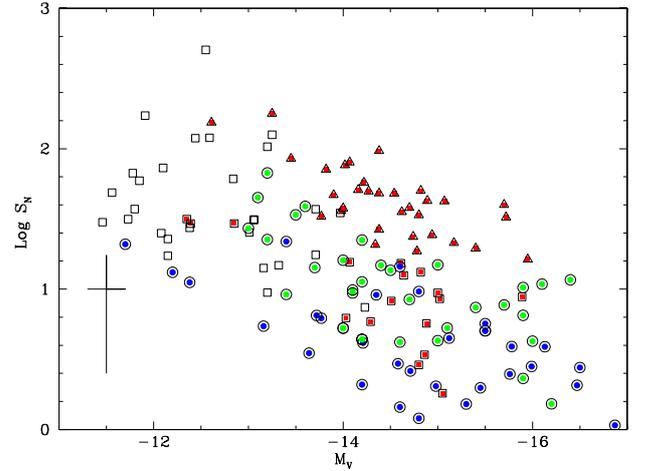}
    \caption{Globular cluster system log specific frequency versus host UDG galaxy absolute magnitude. Coma UDGs are shown by red squares and red triangles (if they have overly massive halos). Coma LSB dwarfs are shown by open squares, Coma early-type dwarfs by green circles and local dwarfs by blue circles.
    The typical uncertainty for Coma UDGs is shown on the left side. 
    While following the general trend, Coma UDGs tend to have higher S$_N$ values than comparable dwarf galaxies. 
     }
    \label{fig:mv}
\end{figure}

In Fig.~\ref{fig:mv} we show S$_N$ vs host galaxy absolute magnitude for the Coma UDGs and dwarf comparison samples.  
The well-known trend of increasing S$_N$ with decreasing luminosity for low mass galaxies can be seen (at higher masses than shown, S$_N$ increases with increasing luminosity forming a U-shaped distribution; \citet{2005ApJ...635L.137F}).
The UDGs follow the general trend seen for the other dwarf galaxies however, along with the LSB dwarfs, they reach significantly higher specific frequencies for a given host galaxy luminosity. These are the same UDGs as the overly massive halo UDGs identified in section 6.3.  
Similar trends were seen by 
\citet{2018ApJ...862...82L}. 
%The lower luminosity LSB dwarfs continue the trend to higher specific frequencies. 

Figures 3 to 6 all suggest that some UDGs have GC systems that are similar in richness to those of other dwarf galaxies. 
However, in the Coma cluster a significant fraction (i.e up to 2/3) of our UDGs reveal GC systems that are much richer than those seen in dwarf galaxies of comparable luminosity or stellar mass. The distribution appears to be continuous without an obvious demarcation (which could be masked by large uncertainties) between the relatively GC-poor and GC-rich UDGs. 

%If we make an arbitrary cut at S$_N$ $>$ 100 (which may select for those UDGs with the most massive halos), we include a dozen UDGs. Only one  early-type dwarf galaxy is similarly GC-rich (see Fig.~\ref{fig:sn}). The GC-rich UDGs tend to be of lower luminosity than the typical UDG, following the general trend of increasing S$_N$ in lower luminosity low mass galaxies. 

In order to better understand why some UDGs are exceptional in terms of their GC richness (and by extension halo mass), we next explore whether S$_N$ depends on other host UDG properties, i.e. size, colour or surface brightness. 

\subsection{GC Specific Fequency vs Size}

Using a sample of UDGs,   
\citet{2018ApJ...862...82L} claimed a weak trend between S$_N$ and host galaxy half-light radius R$_e$ for the more luminous UDGs in their sample and no trend for the least luminous UDGs. Including LSB dwarfs with sizes 0.7 $<$ R$_e$ $<$ 1.5 kpc, \citet{2018MNRAS.475.4235A} hinted at the opposite trend with the most GC-rich galaxies having smaller sizes on average. 

\begin{figure}
	\includegraphics[width=7.1cm, angle=-90]{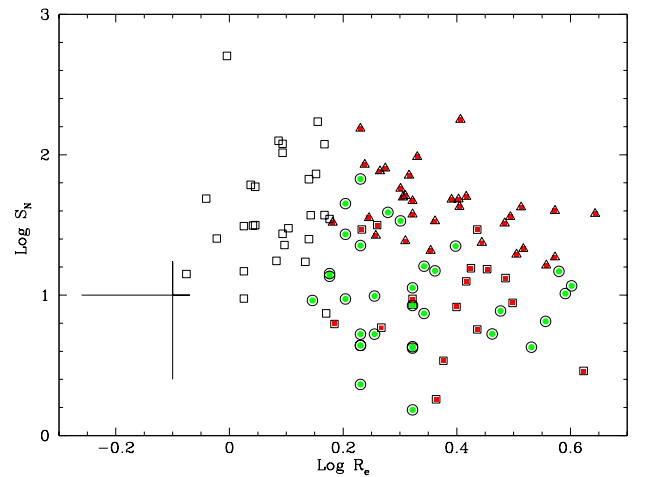}
    \caption{Globular cluster system log specific frequency versus host UDG galaxy half-light radius. Coma UDGs are shown by red squares and red triangles (if they have overly massive halos). LSB dwarfs are shown by open squares and early-type dwarfs by green circles. 
    The typical uncertainty for Coma UDGs is shown on the left side. 
    %The traditional definition of a UDG is R$_e$ $\ge$ 1.5 kpc. 
    There is a weak inverse trend between GC specific frequency and the size of the host galaxy for massive halo UDGs.
}
    \label{fig:size}
\end{figure}

In Fig.~\ref{fig:size} we show GC specific frequency vs half-light radius for Coma UDGs, LSBs and early-type dwarfs. 
Examining the UDGs with massive and normal halos separately, we find no trend for the normal halo UDGs and only a weak inverse trend (at the 2$\sigma$ level) for the massive halo UDGs to have  
higher GC specific frequencies in smaller UDGs. 
%(as might be expected given the trends of S$_N$ with magnitude and of magnitude with size). 
We thus support the general trend found by \citet{2018MNRAS.475.4235A}, albeit not at a statistically significant level and for massive halo UDGs only.

\subsection{GC Specific Frequency vs Colour}

\begin{figure}
	\includegraphics[width=7.1cm, angle=-90]{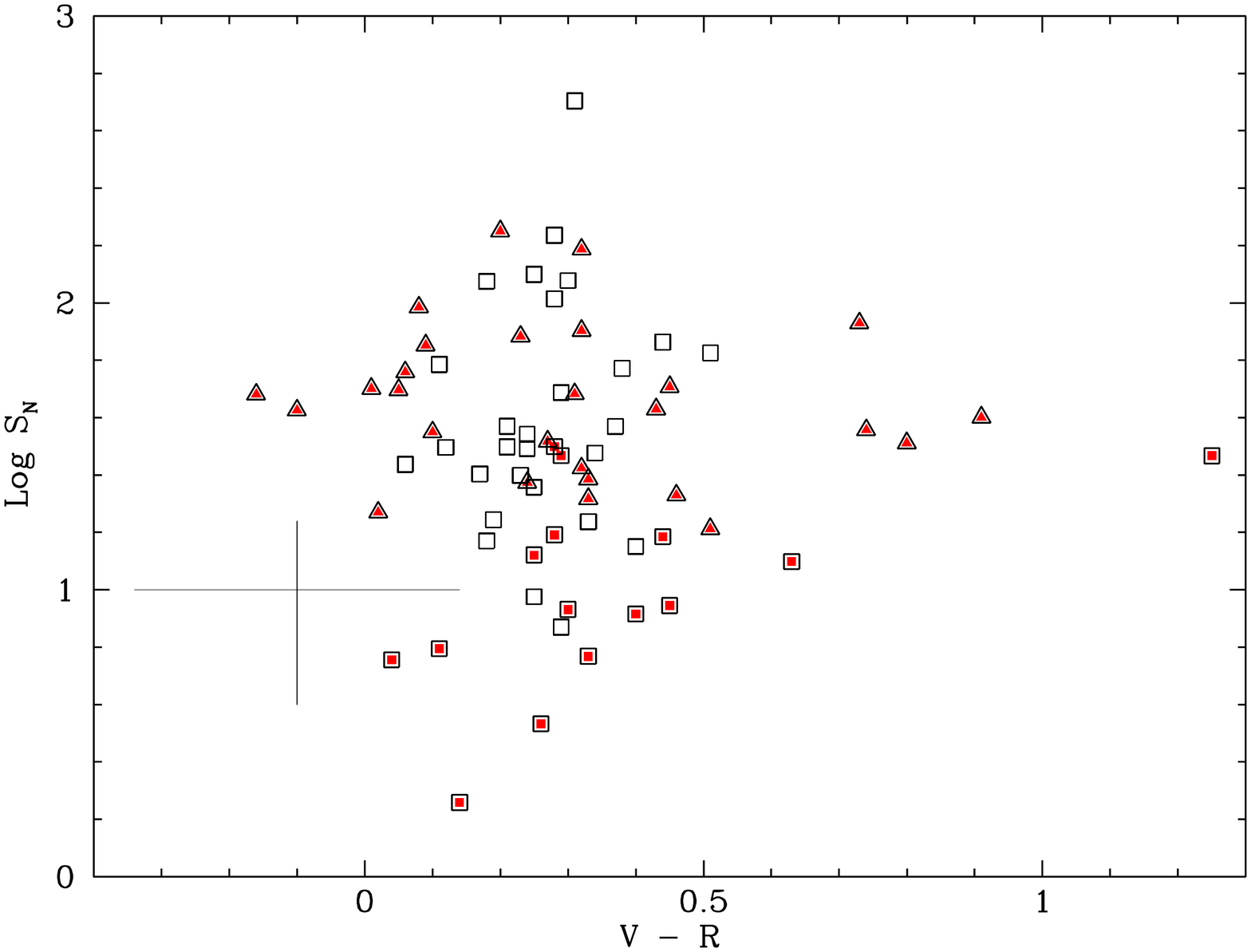}
    \caption{Globular cluster system specific frequency versus host galaxy V--R colour. Coma UDGs are shown by red squares and red triangles (if they have overly massive halos). Coma LSB dwarfs are shown by open squares. 
    The typical uncertainty for Coma UDGs is shown on the left side. 
    There is a trend between GC specific frequency and host galaxy V--R colour for UDGs with normal halos. 
}
    \label{fig:col}
\end{figure}

In Fig.~\ref{fig:col} we show the specific frequency vs host galaxy colour for our UDG sample. 
Examining the UDGs with massive and normal halos separately, we find no trend for the massive halo UDGs. However, the normal halo UDGs reveal a strong trend. A linear best-fit to the data gives:\\
log S$_N$ = 0.61($\pm$0.19) V-R + 0.77($\pm$0.07)\\

Thus we find the redder normal halo UDGs reveal higher GC  specific frequencies on average, whereas 
%We also note an absence of red, low S$_N$ UDGs and that the reddest galaxy shown, DF44 with V--R = 0.91, has a large number of GCs (76) but a typical S$_N$ value. 
the Coma LSB dwarfs 
reveal no clear trend with host galaxy colour.
%As already noted, the two bluest UDGs (DF40 and DF41) are consistent with having no GCs. However several other UDGs also have no GCs and they reveal typical UDG colours of V--R $\sim$ 0.5. 
%At the other extreme, the reddest UDG (Yagi99) has an extremely high S$_N$ value. The next few reddest, however, have fairly typical S$_N$ values.  
%of 355 $\pm$ 276 (only Yagi433 with a typical V--R colour has a higher S$_N$ value of 575 $\pm$ 154). 
%We also note the well-studied UDGs DF42 and DF44, which are both red (V--R = 0.70 and 0.67 respectively) but fairly unremarkable in terms of specific frequency for a UDG, i.e. 25 $\pm$ 19 and S$_N$ = 69 $\pm$ 16 respectively. 
Although we do not find UDG V--R colour  to be a strong characteristic of GC specific frequency for the massive halo UDGs,  future work should re-examine this issue using a more sensitive baseline colour, e.g. combining optical and near-IR filters. We also note that our sample of cluster UDGs lacks the very blue UDGs commonly found in the field.

%One limitation of $S_N$ is that it normalises the GC count by the host galaxy V band luminosity which can be affected by recent star formation. A better approach is to examine the number of GCs per 
%unit stellar mass of the host galaxy (which we call GC richness). 
%In Fig.~\ref{fig:col} we show GC richness vs host galaxy colour for our UDG sample.  As we use a fixed mean GC mass, this is equivalent to the number of GCs per stellar mass. Using stellar mass rather than V band luminosity,  we can now see a mild trend of GC richness with V--R colour. So red (i.e. older and/or more metal-rich) UDGs have, on average, fewer GCs per galaxy stellar mass. The well-known UDG DF44 with $N_{GC}$ = 76 $\pm$ 18 and a red colour (V--R = 0.7) is somewhat of an exception to this trend. 
%However, we caution that much of this trend can be explained by the dependence of stellar mass on galaxy colour (see section 3). 
%We conclude, that if an intrinsic trend exists between GC richness and host galaxy colour it is very weak. 

\subsection{GC Specific Frequency vs Surface Brightness}

\citet{2018ApJ...862...82L} found a weak trend between S$_N$ and host galaxy surface brightness, with higher S$_N$ associated with fainter surface brightnesses. 
%However, this effect was only seen comparing UDGs to early-type dwarf galaxies. 
%Within their sample of UDGs only there was no discernible trend between S$_N$ and surface brightness. 

\begin{figure}
	\includegraphics[width=7.1cm, angle=-90]{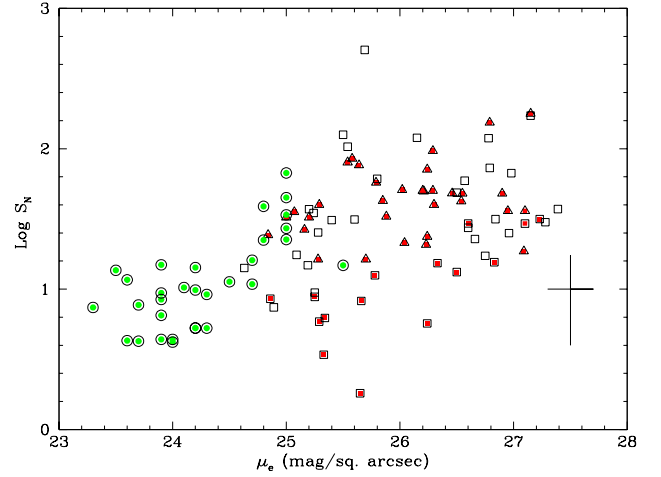}
    \caption{Globular cluster system specific frequency versus host galaxy mean R band surface brightness. Coma UDGs are shown by red squares and red triangles (if they have overly massive halos). Coma LSB dwarfs are shown by open squares. 
    The typical uncertainty for Coma UDGs is shown on the right side. 
    There is a trend between GC specific frequency and host galaxy surface brightness for UDGs with both normal and massive halos.
}
    \label{fig:sb}
\end{figure}

In Fig.~\ref{fig:sb} we show S$_N$ vs host galaxy surface brightness for Coma UDGs, LSB and early-type dwarf samples.
Examining the UDGs with massive and normal halos separately, we find strong trends for both the massive halo and normal halo UDGs. 
Linear best-fits to the data give:\\
log S$_N$ = 0.31($\pm$0.06) $\mu_e$ -- 1.8($\pm$1.6), \\
log S$_N$ = 0.34($\pm$0.08) $\mu_e$ -- 7.9($\pm$2.1)\\
for massive and normal halo UDGs respectively. 
These trends are 
in the same sense as found by \citet{2018ApJ...862...82L}, i.e.  
for fainter surface brightness UDGs to host richer GC systems.

%However we note that that the dozen highest S$_N$ UDGs have surface brightnesses that include essentially the full dynamic range of values. 
%We caution the reader that the weak trend we see with surface brightness could be an observational effect: it being easier to detect GCs in more diffuse galaxies.
%of $\mu_e > 25.4$ mag per sq. arcsec. 
%Thus, with a larger sample of UDGs, we support the claims of \citet{2018ApJ...862...82L}. 

\section{Discussion}

We have investigated the GC systems of Coma UDGs, in comparison with other dwarf galaxy samples, using two measures of GC richness. In the first approach we examined the number of GCs associated with each UDG, and directly related quantity of GC system mass. We find that several Coma UDGs have extremely rich GC systems, relative to other Coma dwarf galaxies, and that their GC systems may contain a significant fraction (e.g.  $\sim$10\%) of their total stellar mass. Assuming that the relation between GC number and halo mass holds for UDGs, the majority have  overly massive halos. 
In particular, we find 65\% of 
Coma UDGs to be more massive than expected compared to the stellar mass-halo mass relation for normal galaxies. These massive halo UDG  
may be considered `failed galaxies', i.e. they lie in  substantial dark matter halos but have failed to 
form many stars. These massive halo UDGs represent a key challenge for contemporary  galaxy formation simulations. The remaining 35\% are consistent with lower mass, or dwarf-like, halos.

Our second approach examined the traditional GC specific frequency (S$_N$), which normalises the GC number by the V band luminosity of the host galaxy. 
We found that the highest S$_N$ Coma UDGs have 
the inter-related properties of having low luminosities, small sizes, and faint surface brightnesses.  For the normal halo UDGs there is also a trend with host galaxy V--R colour, i.e. the UDGs with redder colours tend to host richer GC systems. 

%However these trends are generally shared by our samples of 
%Coma LSB dwarfs (that have smaller sizes than UDGs) 
%and Coma early-type dwarfs (that have higher surface brightnesses than UDGs). %Similar trends are seen for massive halo and `normal' dwarf-like halo UDGs. 
%This 
%indicates that these S$_N$ trends are not unique to UDGs, and suggests some 
%continuity between the different types of dwarf galaxy and the processes determining the richness of their GC systems. 

%galaxy properties, it  
%can also be instructive to look at the extremes of the GC distribution in UDGs, i.e 
%the most GC-rich and those UDGs without any GCs after statistical background subtraction. The richest GC systems tend to be those hosted in UDGs which are of lower luminosity/stellar mass and have slightly fainter surface brightness on average. The UDGs without GCs tend to have large sizes for their luminosity but otherwise have fairly typical properties for UDGs. Otherwise there is little, or no, trend between UDG size, colour or surface brightness with GC richness (and presumably halo mass). 

The origin(s) of UDGs remains a matter of considerable  debate. On the one 
hand, theoretical models have shown that very diffuse dwarfs can arise through internal feedback and/or tidal
effects (\citet{2017MNRAS.466L...1D}; \citet{2015MNRAS.452..937Y}; \citet{2019MNRAS.485..382C};
\citet{2019MNRAS.tmp.1490J}; 
%\citet{2018arXiv181110607J}; 
\citet{2019MNRAS.490.5182L};
\citet{2019MNRAS.485..796M}). 
%\citet{2019arXiv190406356L}). 
On the other hand, some UDGs exhibit peculiar properties that challenge these models, such as their extreme chemical abundance and high dynamical mass within the half-light radius 
(e.g. \citet{2019MNRAS.484.3425M}; 
\citet{2016ApJ...819L..20B}; 
\citet{2018ApJ...856L..31T}; 
\citet{2019ApJ...880...91V}). In this work we have highlighted other properties which challenge these models, namely the high mass halos inferred for those UDGs with rich GC systems. 
% and a paucity of cold gas in some field UDGs \citep{2017A&A...601L..10P}. 

A possible bridge between these different perspectives is that there exist two basic types of UDG,  with similar sizes and luminosities, but distinct  origins. The range in stellar population properties of UDGs found to date lends support to this idea.  
We suggest that the first type consists of ordinary but very low surface brightness large-sized dwarfs. It is these `puffed up' dwarfs that have been successfully reproduced by some recent simulations and they largely form a continuum with other normal dwarf galaxies. The second type is an exotic class of failed galaxy whose halo masses are not reproduced by any current galaxy simulation. 

%If one considers UDGs as pure stellar halos, i.e. with no bulge or disk components \citep{2016ApJ...822L..31P}, then other distinguishing characteristics are also implied. These galaxies, 

Failed galaxies can identified in deep imaging 
by their rich  GC systems, and should have stars with GC-like stellar populations, i.e.  old ages, low  metallicities, and enhanced alpha-element abundances. They should also be more dark matter dominated than normal dwarfs of a similar luminosity, having
missed out on $\sim$10 Gyr of evolution converting gas into stars. The main limitation in testing this concept is a minimal overlap so far between the sample of UDGs with well-studied
GC systems, and those with measurements of their internal dynamics and stellar populations. 
%To date, there is perhaps only one case with all three properties are well established, i.e. DF44 in the Coma cluster. Its stellar populations indicate an early and rapid star formation, and it hosts a large GC system of 76 GCs (\citet{2017ApJ...844L..11V};  Villaume et al. 2019, in prep).

A number of recent works have suggested that star formation at early epochs is dominated by GCs, e.g.  \citet{2018ApJ...869..119E}; \citet{2018MNRAS.479..332B}. 
In the failed galaxy scenario, we would expect 
high S$_N$ to be associated with old, metal-poor UDGs with enhanced [Mg/Fe] (and a high velocity dispersion indicating a massive halo). 
Conversely, low S$_N$ values would be associated with younger ages, higher metallicities, lower [Mg/Fe] (and a low stellar velocity dispersion indicating a dwarf-like dark matter halo). 

In the following we propose a simple mixture model for the two classes of UDG. In this approach UDG  properties vary smoothly from one class to another.
We start by calibrating our simple toy model using the current literature of stellar populations in Coma cluster UDGs. We collect age, total metallicity [Z/H] and alpha element abundance [Mg/Fe] when available from  the works of \citet{2018MNRAS.479.4891F} which includes 7 UDGs, \citet{2018ApJ...859...37G} which gives ages and Iron metallicity [Fe/H] (we assume [Z/H] = [Fe/H] + 0.3) for 3 UDGs, and  \citet{2018MNRAS.478.2034R} which includes data for 5 UDGs (we note that their quoted uncertainties on [Z/H] seem unrealistically small, and we assume $\pm$0.4 for their [Mg/Fe] values). Age and [Fe/H] measurements for 2 Coma UDGs are taken from \citet{2019ApJ...884...79C}
%\citet{2019arXiv190105489C}  
(they also give stellar populations for 6 LSB galaxies that do not meet the standard UDG definition of size and surface brightness, which we do not include here). 
%Recently, \citet{2019A&A...625A..77F}
%obtained age, total metallicity and alpha %element abundance for UDG DF2 in the NGC 1052 %group. 

\begin{figure}
	\includegraphics[width=7.1cm, angle=-90]{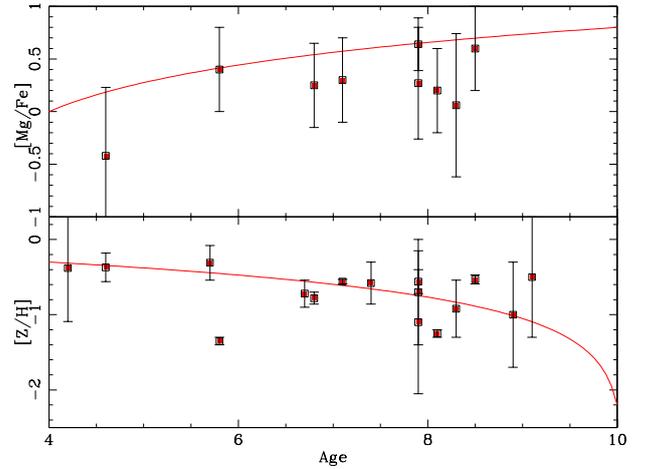}
    \caption{Age--metallicity and age-[Mg/Fe] relations for Coma UDGs with measured stellar populations. The data shown come from three literature studies of Coma UDGs. The typical uncertainty in age is around $\pm$3 Gyr (see Table 1). The red line shows a simple toy model to represent the data (see text for details).
}
    \label{fig:amr}
\end{figure}

In Fig.~\ref{fig:amr} we show the stellar population data for Coma UDGs in terms of [Z/H] and [Mg/Fe] vs age. The plot also shows our simple toy model. We assume GC-like parameters for the stellar populations of a failed galaxy, i.e. [Z/H] = --2.2 and [Mg/Fe] = +0.8 for an age = 10 Gyr. This assumes that the failed galaxy formed its stars rapidly at early times with low metallicity. 
We also assume that a normal dwarf galaxy 
%of stellar mass $\sim10^8 M_{\odot}$
has [Z/H] = --0.3, [Mg/Fe] = 0.0 and an age of 4 Gyr  (see \citet{2009MNRAS.392.1265S} 
for a study of Coma dwarf galaxies). 
The model curves represent a smooth transition from a pure failed galaxy to a fully  
normal dwarf galaxy. The comparison with the data is by no means perfect, but it provides a reasonable representation on average and it can be refined as more data are published.

We further assume in our toy model that a pure failed galaxy UDG forms a significant number of metal-poor GCs, along with its field stars, at early times. 
We note that when the colours of GCs have been measured, they are similar to those of the underlying field stars in the host UDG, e.g. \citet{2016ApJ...819L..20B}. 
Thus a pure failed galaxy, with no subsequent field star formation, would have a very large GC-to-stellar mass ratio, i.e. a very large S$_N$ value. Here we assume S$_N$ = 50, which is near the upper limit observed for Coma UDGs. 

It  
is interesting to compare this GC specific frequency with that of  the halo of our own Milky Way galaxy (given that UDGs appear to lack clear bulge or disk components). 
Indeed the formation of metal-poor GCs at early times may be a universal process \citep{2018RSPSA.47470616F}. 
Similar to the failed galaxy scenario for UDGs, the halo of our Galaxy formed at early times and is dominated by old metal-poor stars. 

To first order, the Milky Way contains around 100 metal-poor GCs today. 
The total stellar mass of the Galactic halo is 5.5($\pm$1.5)$\times$10$^8$ M$_{\odot}$
%, with 2.5($\pm$0.5) $\times$ 10$^8$ M$_{\odot}$ in halo substructures
\citep{2016ARA&A..54..529B}.
%Removing the accreted component, and a
Assuming M/L$_V$ = 1.4 gives a halo luminosity of 3.9$\times$10$^8$ L$_{\odot}$ or M$_V$ = --16.65. %(i.e. comparable to the most luminous Coma cluster UDG). 
The resulting GC specific frequency for the Milky Way halo today (with N$_{GC}$ = 100) is S$_N$ = 22. However, the Galactic environment has destroyed many low-mass GCs over time. If we take the 3\% of all halo stars (or 1.65$\times$10$^7$ M$_{\odot}$) that have signatures of being formed in GCs 
\citep{2011A&A...534A.136M}
and assume this all comes from disrupted GCs with an average GC mass of 10$^5$ M$_{\odot}$, then the halo GC population rises to 265 and S$_N$ = 58. 
%The original number of GCs in the halo may have been even higher than this estimate (see \citet{2018RSPSA.47470616F}). 
 Thus the GC specific frequency of the Milky Way's halo is  comparable to those of the  GC-rich UDGs in the Coma cluster.

For normal dwarfs we assume, on average, S$_N$ = 1 (normal dwarfs range from many to no GCs; see Fig.~\ref{fig:num}). In reality, both normal dwarfs and failed galaxies would have a stochastic variation in S$_N$ values. To summarise, 
a failed galaxy is assigned 
an age of 10 Gyr, [Z/H] = --2.2, [Mg/Fe] = +0.8 and S$_N$ = 50, while a normal dwarf galaxy has an age of 4 Gyr, [Z/H] = --0.3, [Mg/Fe] = 0.0 and S$_N$ =1. Our toy model transitions these properties smoothly from one extreme galaxy to the other.

\begin{figure}
	\includegraphics[width=7.1cm, angle=-90]{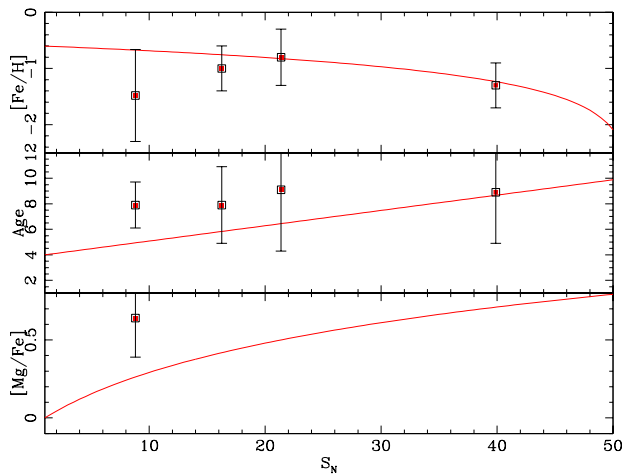}
    \caption{[Mg/Fe], age and [Fe/H] variation with specific frequency. 
    The data shown come from three literature studies of Coma UDGs (see text for details). The red line shows the same simple toy model for UDGs from Fig.~\ref{fig:amr} (assuming high S$_N$ UDGs are failed galaxies with GC-like stellar populations and low S$_N$ are  normal dwarf galaxies). The typical uncertainity in S$_N$ is $\pm$10 (see Table 1). }
    \label{fig:model}
\end{figure}

\begin{table}
	\centering
	\caption{Coma Ultra Diffuse Galaxies with GCs and measured stellar population  properties.}
	\label{tab:example_table}
	\begin{tabular}{lccccr} % four columns, alignment for each
		\hline
		UDG &  S$_N$ & Age & [Fe/H] & [Mg/Fe] & Ref.\\
		\hline
		DF7 & 16$\pm$10 &  7.9$\pm$3.1 & -1.0$\pm$0.4 & --  & G18\\
		DF17 & 21$\pm$9 &  9.1$\pm$4.7& -0.8$\pm$0.5 & -- & G18\\
		DF26 & 9$\pm$9 &  7.9$\pm$1.8 & -1.5$\pm$0.8 & 0.64$\pm$0.25 & FM18\\
		DF44 & 40$\pm$9 &  8.9$\pm$3.8 & -1.3$\pm$0.4 & -- & G18\\
%		Yagi407 & 8.7$\pm$6.3 & 3.0$\pm$0.5 & -0.31$\pm$0.14 & --&  C19\\
		\hline
	\end{tabular}
\end{table}

Unfortunately, only four UDGs have both their stellar populations and GC systems studied to date. We list their stellar population and GC counts (and uncertainty on the counts) in Table 1. 
%These five all have S$_N$ values close to the mean for UDGs. 
%For all but one UDG (Yagi407 from 
%\citet{2019arXiv190105489C}) the host galaxies 
These UDGs are all old and metal-poor, with relatively high S$_N$ values. 
Only one UDG (DF26) has a published alpha element abundance ratio from \citet{2018MNRAS.479.4891F}. 
%We suspect the \citet{2018ApJ...862...82L} value (16$\pm$17) is an underestimate. 
%Here we include an unpublished value from Villaume et al. (2019, in prep) for DF44.
The literature values from Table 1, and our toy model, are shown in Fig.~\ref{fig:model}. Again a comparison between the data and our toy model is not perfect but neither can it be ruled out with confidence. 
The highest S$_N$ galaxy is DF44. This UDG has been well-studied and has a massive halo of $\sim$10$^{11}$ M$_{\odot}$ \citep{2019ApJ...880...91V}, 
making it a good candidate for a failed galaxy.

%Only one UDG (DF26) has a published alpha element abundance ratio from \citet{2018MNRAS.479.4891F}. 
We note that the GC counts for DF26 may be particularly uncertain as it is located near the cluster core and GCs may have been tidally stripped away leaving an abnormally GC-poor galaxy. 
%-- such regions have high levels of surrounding GCs and we suspect its GC count is underestimated. 
%If this is the case then it would move to the left in the plot. 
Measurements of [Mg/Fe] in more UDGs potentially provides a strong point of discrimination with those of normal dwarfs.   Future studies should concentrate on obtaining the stellar populations for UDGs, particularly the [Mg/Fe] ratio, in order to better test and refine this model. 

In this work we focused on UDGs in the Coma cluster, finding many to be GC-rich and by inference occupying overly massive halos. 
While some UDGs in the Virgo cluster are also known to be GC-rich, e.g. \citet{2016ApJ...819L..20B} and  \citet{2018ApJ...856L..31T}), UDGs in the Fornax cluster/group appear to have more modest GC systems \citep{2019MNRAS.484.4865P}. 
In the latter study they measured a typical GC specific frequency of S$_N$ $\sim$ 10 for their dozen Fornax UDGs, with none of them reaching the high S$_N$ levels found in some Coma UDGs. In our toy model, we would predict these low S$_N$ Fornax UDGs to have stars with intermediate ages, [Fe/H] $\sim$ --1 and near solar [Mg/Fe] values.

%Given the age-metallicity relation for our toy model (Fig.~\ref{fig:amr}) we can predict the trend of V--R with S$_N$ from a single stellar population model. Using \citet{1996ApJS..106..307V} (assuming a Kroupa IMF and the baseline alpha-element abundance) the V--R colour ranges from $\sim$0.3 for a pure failed galaxy of very old age and low metallicity to $\sim$0.5 for an intermediate age, more metal-rich galaxy
%(only very metal-rich, old galaxies with some dust would reach V--R $\ge$ 0.6). This suggests a very weak anti-correlation between V--R and S$_N$. 
%For example, between 12 and 8 Gyr we expect the colour to change from V--R = 0.4 to 0.5 as S$_N$ goes from 100 to $\sim$50. 
%Thus we do not expect V--R to correlate strongly with S$_N$ (as seen in  Fig.~\ref{fig:col}). The age-metallicity degeneracy means that a single optical colour is a poor diagnostic of the differences between a failed galaxy and a normal dwarf galaxy. 

\section{Conclusions}

%\citet{2018ApJ...862...82L} found a trend for 
%UDGs near the centre of the Coma cluster to have richer GC systems. 
%and a weak trend with galaxy surface brightness. 
%They also claimed a weak trend for more luminous UDGs with large sizes to have richer GC systems. 

Using an enlarged sample and deeper imaging of UDGs in the Coma cluster we study their globular cluster (GC) systems. We find that Coma UDGs are on average 
overabundant in GCs compared to  Coma early-type dwarfs and local dwarfs 
of a comparable host galaxy stellar mass. However, Coma LSB dwarfs (with sizes 0.7 $<$ R$_e$ $<$ 1.5 kpc) are similarly over-abundant in GCs. 
In some extreme cases, the mass contained in the GC system is around 10\% of the total stellar mass of its host UDG. Coma UDGs can have specific frequencies up to S$_N$ $\sim$ 100.  
The distribution of GC specific frequency, and of GC number counts, appears to be rather continuous (although the measured uncertainties are large, potentially 
masking the presence of clear GC-rich and GC-poor subclasses).

Like normal dwarf galaxies in Coma, UDGs reveal a trend for 
higher GC specific frequencies to be found in lower mass hosts. 
Higher GC specific frequencies are also found in 
lower surface brightness UDGs and in 
redder normal halo UDGs.  
There is a weak trend for higher GC specific frequencies in 
smaller sized massive halo UDGs.

From the scaling of GC system numbers to halo mass, we infer that around 2/3 of our Coma UDGs have massive halos indicative of failed  galaxies. Such high halo masses ($\sim10^{11}$~ M$_{\odot}$) 
are not reproduced by current galaxy simulations, which have had some success in finding UDGs with dwarf-like halos.

We present a simple mixture model as a first step to help interpret the large fraction of Coma UDGs with extraordinarily rich GC systems (which imply massive halos and high GC specific frequencies). We propose that such GC-rich UDGs formed both their GCs and field stars rapidly at early times. Thus the stellar populations of both GCs and field stars would be metal-poor with enhanced alpha element ratios and with very old ages. If these UDGs failed to form any additional field stars (e.g. due to mass loss, gas stripping etc) without disrupting their GC systems, it would lead to very high GC specific frequencies, as observed. This `failed galaxy' scenario may thus explain the GC-rich, dark matter-dominated UDGs.  The GC-poor UDGs would be more akin to normal dwarf galaxies that have been puffed-up to larger sizes by internal feedback processes or tidal interactions.  
%ongoing star formation over %cosmic time that raises their %mean stellar metallicities %while reducing their GC %specific frequencies to %values more typical of dwarf %galaxies. 
%In this toy model we do not expect to find a detectable trend between UDG V--R colour and GC richness. 
Future observational studies should concentrate on obtaining additional stellar population (especially alpha elements) and dynamical properties for UDGs with extreme (high and low) GC numbers, in order to better test and refine this model. Future simulations of UDGs need to incorporate realistic GC systems.

\section*{Acknowledgements}

DAF thanks the ARC for financial support via DP160101608. JPB and AJR acknowledge financial support from AST-1616598, AST-1518294 and 
AST-1616710. AJR was supported by the Research Corporation for Science Advancement as a Cottrell Scholar. We thank J. Gannon for useful comments. NA is supported by the Brain Pool Program, which is
funded by the Ministry of Science and ICT through the National Research 
Foundation of Korea (2018H1D3A2000902). We thank the referee for several useful comments.

%%%%%%%%%%%%%%%%%%%%%%%%%%%%%%%%%%%%%%%%%%%%%%%%%%



\bibliographystyle{mnras}
\bibliography{bbl} % if your bibtex file is called example.bib

%\begin{thebibliography}{99}
%Author A.~N., 2013, Journal of %Improbable Astronomy, 1, 1
%\end{thebibliography}

%%%%%%%%%%%%%%%%%%%%%%%%%%%%%%%%%%%%%%%%%%%%%%%%%%

%%%%%%%%%%%%%%%%% APPENDICES %%%%%%%%%%%%%%%%%%%%%

\appendix

    \section{Globular cluster numbers}
    
Here we list ultra diffuse galaxies in the Coma 
cluster GCs used in this work. Tables A1 lists those with Dargonfly IDs and Table A2 with Yagi IDs. Table A3 lists UDGs  from \citet{2018ApJ...862...82L} that are covered by the photometry of Alabi et al. (2019, in prep.). 
%The photometric properties of the UDG host galaxies come from Alabi et al. (2019, in prep.)

\begin{table}
	\centering
	\caption{The number of globular clusters in Coma cluster ultra diffuse galaxies with both Yagi and Dragonfly IDs. 
	}
	\label{tab:example_table}
	\begin{tabular}{llccc} % four columns, alignment for each
		\hline
		ID1 & ID2 & M$_V$ & N$_{GC}$ & $\pm$\\
		\hline
yagi13 &DFX1    &   -15.72      &    63  &        17\\
yagi851 &DF9    &   -13.98      &  14.1  &      13.7\\
yagi194 &DF8    &   -14.82      &  42.6  &      23.9\\
yagi680 &DF7    &   -15.95      &  39.1  &      23.8\\
yagi853 &DF6    &   -14.54      &  31.4  &      22.6\\
yagi8 &DF46     &  -14.47       &    0   &     15.7\\
yagi11 &DF44    &    -15.70      &    76  &        18\\
yagi14 &DF42    &   -14.64      &     9  &         7\\
yagi501& DF41   &    -15.14     &   -1.9 &       14.3\\
yagi507& DF40   &    -14.95     &   -7.1 &       10.3\\
yagi581& DF39   &    -14.82     &   11.2 &       17.6\\
yagi762& DF36   &    -14.27     &   25.4 &       15.7\\
yagi37 &DF35    &   -14.07      &   6.6  &      14.4\\
yagi782& DF32   &    -14.88     &    5.1 &       13.1\\
yagi739& DF31   &    -14.38     &   27.2 &         17\\
yagi577& DF29   &    -15.07     &   45.1 &       21.5\\
yagi93 &DF26    &   -15.89      &    20  &      20.7\\
yagi285& DF25   &    -14.61     &   10.7 &       24.3\\
yagi364& DF23   &    -14.89     &   38.5 &       19.2\\
yagi86 &DF20    &   -13.53      &  -0.1  &       9.9\\
yagi484& DF2    &    -14.70      &     0  &       7.8\\
yagi347& DF18   &    -13.77     &   10.6 &       19.1\\
yagi165& DF17   &    -15.17     &     25 &         11\\
yagi660& DF15   &    -14.78     &   15.2 &       22.6\\
yagi653& DF14   &    -14.38     &   54.6 &       24.2\\
yagi163& DF13   &    -14.34     &   11.3 &       10.5\\
yagi416& DF12   &    -14.16     &   23.5 &       17.4\\
yagi459& DF10   &    -15.05     &    1.9 &       10.9\\
yagi486& DF1    &   -14.74      &  18.6  &      13.9\\
		\hline
	\end{tabular}
\end{table}

\begin{table}
	\centering
	\caption{The number of globular clusters in Coma cluster ultra diffuse galaxies with only Yagi IDs.
	}
	\label{tab:example_table}
	\begin{tabular}{llccc} % four columns, alignment for each
		\hline
		ID1 & ID2 & M$_V$ & N$_{GC}$ & $\pm$ \\
\hline
yagi534 &--   &    -14.22     &     28     &     11\\
yagi436 &--   &    -14.07     &     34     &     13\\
yagi425 &--   &    -13.82     &     24     &     11\\
yagi424 &--   &    -12.61     &     17     &      9\\
yagi419 &--   &    -14.94     &     23     &     11\\
yagi386 &--   &    -14.86     &      3     &      8\\
yagi367 &--   &    -14.02     &     31     &     12\\
yagi122 &--   &    -13.83     &      0     &      4\\
yagi121 &--   &    -14.62     &     25     &     11\\
yagi112 &--   &    -14.38     &     15     &      9\\
yagi99 &--    &   -13.25      &  35.5      & 27.34\\
yagi91 &--    &   -12.08      &   1.7      &  1.25\\
yagi89 &--    &   -13.71      &  11.3      &  8.44\\
yagi85 &--    &    -13.20     &    1.8     &   1.25\\
yagi438 & --  &     -12.85    &    4.05    &    3.01\\
yagi437 &--   &    -13.06     &    5.2     &   3.75\\
yagi435 &--   &    -11.78     &   3.45     &   2.54\\
yagi433 &--   &    -12.55     &  52.95     &  14.18\\
yagi432 &--   &    -13.25     &   25.1     &  12.73\\
yagi427 &--   &    -12.15     &   1.25     &    0.9\\
yagi423 &--   &     -13.20    &    19.7    &    9.22\\
yagi415 &--   &     -12.10    &    5.05    &    3.79\\
yagi412 &--   &    -12.84     &   8.35     &   5.74\\
yagi410 &--   &    -13.16     &    2.6     &    1.8\\
yagi409 &--   &    -14.51     &   5.25     &   3.71\\
yagi408 &--   &    -11.85     &   3.25     &   2.46\\
yagi407 &--   &    -15.02     &    8.7     &   6.25\\
yagi402 &--   &    -13.01     &   4.05     &   2.93\\
yagi395 &--   &    -12.38     &   2.45     &   1.84\\
yagi391 &--   &    -11.56     &   2.05     &   1.52\\
yagi387 &--   &    -13.97     &   13.5     &   8.83\\
yagi380 &--   &    -12.39     &   2.65     &   1.91\\
yagi374 &--   &    -13.06     &   5.25     &   3.79\\
yagi373 &--   &    -14.29     &   3.05     &   2.07\\
yagi372 &--   &    -13.71     &   5.35     &   3.87\\
yagi331 &--   &    -14.03     &   2.55     &   1.76\\
yagi238 &--   &    -13.32     &   3.15     &   2.23\\
yagi236 &--   &    -12.44     &  11.25     &    7.3\\
yagi118 &--   &    -12.15     &   1.65     &   1.21\\
yagi115 &--   &    -11.91     &     10     &   7.58\\
yagi114 &--   &    -14.23     &   3.65     &   2.54\\
yagi113 &--   &     -11.80    &    1.95    &    1.45\\
yagi108 &--   &    -11.46     &   1.15     &   0.82\\
yagi107 &--   &    -12.35     &   2.75     &   2.07\\
yagi105 &--   &    -11.73     &   1.55     &   1.13\\
yagi104 &--   &    -13.45     &   20.4     &   8.28\\
yagi102 &--   &    -12.59     &     13     &   6.25\\
		\hline
	\end{tabular}
\end{table}

\begin{table}
	\centering
	\caption{The number of globular clusters in Coma cluster ultra diffuse galaxies with photometry from \citet{2018ApJ...862...82L}. 
	}
	\label{tab:example_table}
	\begin{tabular}{llccc} % four columns, alignment for each
		\hline
		ID1 & ID2 & M$_V$ & N$_{GC}$ & $\pm$ \\
\hline
-- &DF4 &-14.5     &   -18     &   36.8\\       
Yagi176 &DF11 & -15.0 & 9.4     &   12.7\\       
-- &DF19 & -14.7 &          28.8 &       23.4\\       
-- &DF22 & -14.0 &          15 &       11.3\\       
-- &DF30 & -15.4 &       28.2 &       18.4\\       
Yagi774 &DF34 & -13.8 &        -2.7 &       12.4    \\   
-- &DF47 & -14.8 &          2.4    &    16.9\\       
yagi358 & -- &-14.8 &         28 &        5.3\\       
yagi370 & -- &-13.9 &        17 &        6.4\\       
		\hline
	\end{tabular}
\end{table}

%%%%%%%%%%%%%%%%%%%%%%%%%%%%%%%%%%%%%%%%%%%%%%%%%%

% Don't change these lines
\bsp	% typesetting comment
\label{lastpage}
\end{document}